\documentclass[12pt]{article}
\usepackage{hepth}
\usepackage{graphicx}

\newcommand{\poly}{\ensuremath{f(r)}}

\newcommand{\coefone}{\ensuremath{C}}
\newcommand{\coeftwo}{\ensuremath{D}}

\begin{document}

\preprint{LMU-ASC 48/08 \\ MPP-2008-116}

\institution{MPI}{Max Planck-Institut f\"ur Physik
  (Werner-Heisenberg-Institut), F\"ohringer Ring 6, \cr 80805
  M\"unchen, Germany}

\institution{LMU}{Arnold Sommerfeld Center for Theoretical Physics, Ludwig-Maximilians-Universit\"at, \cr 
Department f\"ur Physik, Theresienstrasse 37, 80333 M\"unchen, Germany}

\title{Fluid dynamics of R-charged black holes}

\authors{
Johanna~Erdmenger\worksat{\MPI,}\footnote{e-mail: {\tt jke@mppmu.mpg.de}},
Michael~Haack\worksat{\LMU,}\footnote{e-mail: {\tt michael.haack@physik.uni-muenchen.de}},
Matthias~Kaminski\worksat{\MPI,}\footnote{e-mail: {\tt kaminski@mppmu.mpg.de}},
Amos~Yarom\worksat{\LMU,}\footnote{e-mail: {\tt amos.yarom@physik.uni-muenchen.de}}
}

\abstract{We construct electrically charged AdS${}_5$ black hole solutions whose charge, mass and boost-parameters vary slowly with the space-time coordinates. From the perspective of the dual theory, these are equivalent to hydrodynamic configurations with varying chemical potential, temperature and velocity fields. We compute the boundary theory transport coefficients associated with a derivative expansion of the energy momentum tensor and $R$-charge current up to second order.
In particular, for the current we find a first order transport
coefficient associated with the vorticity of the fluid.}

\date{August 2008}

\maketitle

\tableofcontents

\section{Introduction}
The AdS/CFT correspondence has proved to be a useful tool in understanding various aspects of strongly coupled gauge theories. In this work we focus on some developments which allow one to relate the hydrodynamic regime of the gauge theory to black hole solutions in asymptotically AdS${}_5$ backgrounds. The first work in this direction was carried out in  \cite{Policastro:2001yc} where the ratio of the shear viscosity $\eta$ to the entropy density $s$ of the $\mathcal{N}=4$ $SU(N)$ supersymmetric Yang-Mills theory was computed via the Kubo formula. It was found that at strong t' Hooft coupling and in the large $N$ limit
\begin{equation}
\label{E:etaovers}
	\frac{\eta}{s} = \frac{1}{4\pi}.
\end{equation}
This value seems to be universal, and applies to a large class of theories which have a holographic dual \cite{Buchel:2003tz,Kovtun:2008kw,Buchel:2004qq,Son:2006em,Benincasa:2006fu,Brustein:2008cg}, see \cite{Kovtun:2004de} for a review. Finite $N$ corrections to \eqref{E:etaovers} were
considered in \cite{Buchel:2004di,Benincasa:2005qc,Benincasa:2006fu,Brigante:2007nu,Dutta:2008gf,Myers:2008yi,Brustein:2008cg}.
The fact that the black hole background allows to compute hydrodynamic
transport coefficients might be an indication that black holes capture the full hydrodynamic behavior of the boundary theory.
In \cite{Bhattacharyya:2007jc}
an important step in this direction was made: it was shown how to map a hydrodynamic expansion of the
boundary theory to a gradient expansion in the bulk. In principle, this
technique
allows to compute all the transport coefficients
of conformal fluid dynamics (for gauge theories with an AdS dual). Those coefficients which are accessible via linear response theory can also be computed using the Kubo formula. For instance, in 
\cite{Baier:2007ix,Natsuume:2007ty} some of the second order transport coefficients were computed this way.

The method of \cite{Bhattacharyya:2007jc} has been shown to be rather robust and can be applied to black holes in various dimensions \cite{VanRaamsdonk:2008fp,Haack:2008cp}, and to situations where the black hole metric couples to external fields such as the dilaton \cite{Bhattacharyya:2008ji}, implying forced fluid dynamics in the boundary theory. The method involves extending known asymptotically AdS${}_5$ black hole solutions by allowing various parameters of the solution to vary with the space-time coordinates.
In this work we consider charged AdS${}_5$ black holes.
We show that the known Reissner-Nordstr\"om charged black hole solutions can be extended so that their charge, mass and certain boost parameters are slowly varying in the
coordinates transverse to the AdS radial direction (henceforth, the transverse coordinates). In the dual picture, this corresponds to a hydrodynamic limit of the theory where the charged current, energy density, and velocity fields are slowly varying.

In detail, the bulk theory we have in mind is Einstein-Maxwell gravity with a negative cosmological constant and a Chern-Simons term.
This is a consistent truncation of IIB supergravity on AdS${}_5\times S^5$ and is dual to the strongly coupled, planar limit of the $\mathcal{N}=4$ $SU(N)$ supersymmetric-Yang-Mills theory on $\mathbf{R}^{3,1}$ with a non vanishing chemical potential  \cite{Cvetic:1999ne,Chamblin:1999tk}. We will sometimes call this SYM or $\mathcal{N}=4$ theory for short.
More precisely, it is dual to a subsector of the $\mathcal{N}=4$ theory in which a single conserved $U(1)$ current is excited. This is the Noether current associated with the diagonal $U(1)$ of the maximal Abelian subgroup of the $SO(6)$ $R$-symmetry group and it is dual to the bulk $U(1)$ gauge field. More details on this truncation can be found in \cite{Chamblin:1999tk}.

Previous computations of the thermodynamic properties of the SYM fluid with a finite chemical potential can be found in \cite{Chamblin:1999tk,Cvetic:1999ne,Chamblin:1999hg} where the energy density and equation of state have been analyzed. The shear viscosity of this fluid has been computed via the Kubo formula in \cite{Son:2006em,Maeda:2006by,Mas:2006dy}, in \cite{Son:2006em} its heat conductivity was analyzed and in \cite{Natsuume:2007ty} some of the dispersion relations were computed to second order in a small momentum expansion. In \cite{Saremi:2006ep} one can find a related analysis dealing with M2-branes. In this work, we extend these results and compute all second order, linear and non-linear, transport coefficients. 
We also find a first order contribution to the $R$-charge current which was not considered in the literature so far.

The rest of this paper is organized as follows: in the subsequent section we discuss conformal fluid hydrodynamics, set the notation for the rest of this paper and summarize our field theory results. In section \ref{S:RNBH} we review the Reissner-Nordstr\"om AdS${}_5$ black hole solution and rederive the thermodynamic properties of the associated boundary theory. Our main computation, extending the Reissner-Nordstr\"om black hole solution to one with a slowly varying charge, mass and boost parameters is done in section \ref{S:Main}. There, we also explain how to compute the associated transport coefficients. We end with a discussion in section \ref{S:Discussion}. Some of the details of the analysis are relegated to an appendix.

Towards the end of this work we learned about \cite{Banerjee:2008th} which has some overlap with the material presented here.

\section{Conformal fluid dynamics (and summary)}
\label{S:Conformalfd}
Consider the energy momentum tensor $\langle T^{\mu\nu} \rangle$ and a conserved current $\langle J^{\mu} \rangle$ of a conformal theory in $\mathbf{R}^{3,1}$. In the absence of anomalies one has
\begin{equation}
\label{E:Conservation}
	\partial_{\mu} \langle T^{\mu\nu} \rangle = 0\,\quad \partial_{\mu} \langle J^{\mu} \rangle = 0,
\end{equation}
and the energy momentum tensor is traceless
\begin{equation}
\label{E:Trace}
	\langle T^{\mu}_{\mu} \rangle=0.
\end{equation}
In the hydrodynamic approximation, where the mean free path of the theory $\ell_{\rm mfp}$ is smaller than the typical inverse momentum scale, both $\langle T_{\mu\nu} \rangle $ and $\langle J_{\mu} \rangle$ can be expressed in terms of hydrodynamic fields. These are given by the energy density $\epsilon$, the charge density $\rho$ and the velocity field $u^{\mu}$ (which is normalized such that $u^{\mu}u_{\mu}=-1$). In the Landau frame, which we use in the rest of this work, all the hydrodynamic fields are defined relative to the rest frame of a fluid element.
The energy density is given by the time-time component of the energy momentum tensor in the rest frame of the fluid element, the charge density is given by the zero component, $J^0$, of the current in the rest frame of a fluid element, and the velocity of a fluid element is defined by the boost parameter needed to bring that fluid element to its rest frame. In general, one can exchange the energy density and charge density with the temperature $T$ and chemical potential $\mu$, though their explicit functional relation will depend on the specific details of the theory.
For the $\mathcal{N}=4$ theory these relations were computed in \cite{Chamblin:1999tk,Chamblin:1999hg,Son:2006em} and are given in  \eqref{E:epsilon} and \eqref{E:rho} together with \eqref{E:UsefulIDs}. In, for instance, \cite{Erdmenger:2008yj} one can find such relations when flavored matter is introduced.

Working in the hydrodynamic regime, where the momentum scale is smaller than the inverse mean free path, implies that the velocity field, energy density and charge density vary slowly with the space-time coordinates, i.e.\ their derivatives are small. For example, one has
\begin{equation}
	| \partial\epsilon | \ll \epsilon/ \ell_{\rm mfp}\ .
\end{equation}
In this case, we can expand the energy momentum tensor and current in gradients of the hydrodynamic variables. At zero order in such a gradient expansion (meaning a fluid with constant energy density, charge density and moving at a fixed velocity), the only current which can be constructed from $u_{\mu}$, $\epsilon$ and $\rho$ will be proportional to $u_{\mu}$. The only symmetric traceless tensor one can construct must be proportional to $\eta_{\mu\nu}+u_{\mu}u_{\nu}$.
Thus, to leading order in gradients,
\begin{equation}
\label{E:PerfectTandJ}
	\langle T_{\mu\nu} \rangle=
		\frac{\epsilon}{3} \left(4 u_{\mu}u_{\nu}+\eta_{\mu\nu}\right)\ ,\quad
	\langle J_{\nu} \rangle = \rho u_{\nu}\ .
\end{equation}
We denote higher order gradient corrections to the energy momentum tensor and current by $\Pi_{\mu\nu}$ and $\Upsilon_{\mu}$,
\begin{equation}
\label{E:corrections}
	\langle T_{\mu\nu} \rangle =
		\frac{\epsilon}{3} \left(4 u_{\mu}u_{\nu}+\eta_{\mu\nu}\right)+\Pi_{\mu\nu}\ ,\quad
	\langle J_{\nu} \rangle = \rho u_{\nu}+\Upsilon_{\nu}.
\end{equation}
Working in the Landau frame, we find that $u^{\nu}\Upsilon_{\nu}=0$ and $u^{\nu}\Pi_{\mu\nu}=0$.

Following \cite{Baier:2007ix} it is possible to construct the form of the corrections $\Upsilon_{\mu}$ and $\Pi_{\mu\nu}$, order by order in a derivative expansion. Let us start by evaluating all possible contributions to $\Upsilon_{\nu}$ at first order in a gradient expansion. Overall, there are four possible vectors that can be constructed from $\epsilon$, $\rho$ and $u_{\nu}$ which are orthogonal to the velocity field and have one derivative. These are
\begin{align}
\label{E:Vs}
\notag
	V_{\mu}^{1} &= P_{\mu}^{\phantom{\mu}\nu} \partial_{\nu} \epsilon &
	V_{\mu}^{2} &= P_{\mu}^{\phantom{\mu}\nu} \partial_{\nu} \rho &
	V_{\mu}^{3} &= P_{\mu}^{\phantom{\mu}\nu} u^{\alpha}\partial_{\alpha} u_{\nu}\\
	\widetilde{V}_{\mu}^{1} &= \ell_{\mu}\ ,
\end{align}
where we have defined
\begin{equation}
\label{E:defell}
	\ell_{\mu} =\epsilon_{\mu}^{\phantom{\mu}\rho\sigma\tau}u_{\rho}\partial_{\sigma}u_{\tau}\ ,
\end{equation}
and $P_{\mu\nu}$ projects onto the space orthogonal to the velocity field,
\begin{equation}
\label{E:defProjection}
	P_{\mu\nu}=u_{\mu}u_{\nu}+\eta_{\mu\nu}.
\end{equation}
$\ell_{\mu}$ reduces to the curl of the velocity in the local rest frame. In the following, we will always adorn vectors or tensors involving $\ell_{\mu}$ with a tilde.

Since energy conservation \eqref{E:Conservation} implies that
\begin{equation}
	V^1_\nu = -4 \epsilon V^3_\nu,
\end{equation}
we can construct the leading derivative terms in $\Upsilon_\nu$ by using combinations of only $V^1_\nu$, $V^2_\nu$ and $\widetilde{V}^1_\nu$. Further, recall that in a conformal theory, a conserved current $J_{\mu}$ should transform homogeneously under Weyl transformations (we use the conventions of \cite{Baier:2007ix} where $\eta_{\mu\nu} \to e^{-2\omega}\eta_{\mu\nu}$).
$\widetilde{V}^1_{\mu}$ is Weyl invariant while $V^1_\nu$ and $V^2_\nu$ transform inhomogeneously with the inhomogeneous terms given by
\begin{equation}
	\delta V^1_{\mu} = e^{4\omega} P^{\phantom{\mu}\nu}_{\mu} 4 \epsilon \partial_{\nu}\omega\, ,\qquad
	\delta V^2_{\mu} = e^{3\omega} P^{\phantom{\mu}\nu}_{\mu} 3 \rho \partial_{\nu}\omega.
\end{equation}
However, the linear combination
\begin{equation}
	P^{\phantom{\mu}\nu}_{\mu} \partial_{\nu} \frac{\epsilon^3}{\rho^4}
\end{equation}
does transform homogeneously.
One could have guessed this by noting that the weight of $\epsilon$ under Weyl rescalings is 4 and the weight of $\rho$ under Weyl rescalings is 3, so the only Weyl invariant combination is $\epsilon^3/\rho^4$. Switching from energy density and charge density to temperature $T$ and chemical potential $\mu$, we find that
the most general form of $\Upsilon_{\nu}$ at first order in a derivative expansion is
\begin{equation}
\label{E:Upsilonfirst}
	\Upsilon_{\nu} = -\kappa P^{\phantom{\nu}\alpha}_{\nu} \partial_{\alpha} \frac{\mu}{T} + \Omega \ell_{\nu} + \mathcal{O}(\partial^2)\ ,
\end{equation}
where $\kappa=\kappa(\mu,T)$ and $\Omega=\Omega(\mu,T)$ are undetermined first order transport coefficients whose explicit form depends on the theory. Our value for $\kappa$ and $\Omega$ can be found in \eqref{E:kappa} and \eqref{E:omega}

In principle, the transport coefficient $\kappa$ in \eqref{E:Upsilonfirst} can also be calculated via linear response theory. In \cite{Son:2006em} it has been calculated for a different sector of the $\mathcal{N}=4$ theory in which a different $U(1)$ subgroup of the $R$-charge current is excited in addition to some of the scalar fields. The coefficient $\Omega$ in \eqref{E:Upsilonfirst} has been inaccessible so far.\footnote{From the bulk point of view, it is the Chern-Simons term in the action (cf.\ \eqref{E:action} below) that is responsible for having $\Omega \neq 0$.} 

In \cite{LandL} a different type of argument has been used to construct the first order terms in $\Upsilon_{\nu}$: an entropy current was constructed by hand to have a positive semi-definite divergence and from it, the first order corrections to $\Upsilon_{\nu}$ were inferred. The $\Upsilon_{\nu}$ constructed in \cite{LandL} differs from the one in \eqref{E:Upsilonfirst} by the $\ell_{\nu}$ term. It would be interesting to find a corrected form of the entropy current which allows for the $\ell_{\nu}$ term, perhaps along the lines of \cite{Bhattacharyya:2008ji}.

Second order contributions to $\Upsilon^{\nu}$ can be derived using the same arguments as those leading to \eqref{E:Upsilonfirst}. One may construct all possible Weyl-covariant vectors composed of two derivatives which are orthogonal to the velocity field. See for example \cite{Erdmenger:1997gy,Erdmenger:1997wy,Iorio:1996ad,Loganayagam:2008is,Bhattacharyya:2008xc,Bhattacharyya:2008ji,Baier:2007ix} for a more elaborate discussion on the construction of Weyl invariant quantities. In this work five such terms will be relevant,\footnote{Note that all the terms in \eqref{E:Xidef} are of order two in the derivative expansion. Their superscripts are simply a means to enumerate them.}
\begin{align}
\notag
    \Xi^{(1)}_{\nu} &= \sigma_{\nu}^{\phantom{\nu}\alpha}\partial_{\alpha} \frac{\mu}{T}\ ,&
    \Xi^{(2)}_{\nu} &= \omega_{\nu}^{\phantom{\nu}\alpha}\partial_{\alpha} \frac{\mu}{T}\ ,&
    \Xi^{(3)}_{\nu} &= P^{\beta}_{\phantom{\beta}\nu}\partial_{\alpha}\left(\sigma^{\alpha}_{\phantom{\alpha}\beta} b^{-3}\right)\ ,&
    \Xi^{(4)}_{\nu} &= P^{\beta}_{\phantom{\beta}\nu}\partial_{\alpha}\left(\omega^{\alpha}_{\phantom{\alpha}\beta} b^{-1}\right)\\
    \widetilde{\Xi}^{(1)}_{\nu} &= \sigma_{\nu\alpha}\ell^{\alpha}\ ,
\label{E:Xidef}
\end{align}
where we have defined\footnote{In the published version of our paper there was a typographical error in the overall sign of $\omega_{\mu\nu}$. This led to an apparent discrepancy between our results and those of reference \cite{Banerjee:2008th}, mentioned in version 2 of that paper. We thank the authors of \cite{Banerjee:2008th} for pointing out this mismatch. With the current sign conventions, our results are in complete agreement with \cite{Banerjee:2008th}.}
\begin{equation}
\label{E:defsw}
 	\sigma_{\mu\nu} = 2\partial_{\langle \mu}u_{\nu\rangle}\ , \quad
	\omega_{\mu\nu} = \frac{1}{2}P_{\mu}^{\lambda}P_{\nu}^{\sigma}\left(\partial_{\sigma}u_{\lambda} - \partial_{\lambda}u_{\sigma} \right)
\end{equation}
and angular brackets denote a traceless projection onto the space orthogonal to $u_{\mu}$ so that
\begin{equation}
    A_{\langle \mu \nu \rangle}
    = P_{\mu}^{\lambda}P_{\nu}^{\sigma}\frac{1}{2}\left(A_{\lambda\sigma}+A_{\sigma\lambda}\right)
      -\frac{1}{d-1}P_{\mu\nu}P^{\lambda\sigma}A_{\lambda\sigma}
\end{equation}
satisfies $\eta^{\mu\nu} A_{\langle \mu \nu \rangle} = 0$ and $u^{\mu}A_{\langle \mu\nu \rangle} = 0$. 
For the $R$-charge current of the theory at hand, the only non vanishing second order transport coefficients are those associated with the terms in \eqref{E:Xidef}.
Thus, up to second order in a derivative expansion, we may write
\begin{align}
\label{E:GeneralJ}
	\Upsilon_{\nu} =
	-\kappa P_{\nu}^{\alpha}\partial_{\alpha}\frac{\mu}{T} + \Omega \ell_{\nu}+
	\xi_1 \Xi^{(1)}_{\nu}+\xi_2 \Xi^{(2)}_{\nu}+\xi_3 \Xi^{(3)}_{\nu}+\xi_4 \Xi^{(4)}_{\nu}+\tilde{\xi}_1 \widetilde{\Xi}^{(1)}_\nu.
\end{align}

The decomposition of $\Pi_{\mu\nu}$ into Weyl invariant tensors may be carried out in a similar manner \cite{Baier:2007ix}. For the $\mathcal{N}=4$ theory the energy momentum tensor takes the form
\begin{equation}
\label{E:GeneralT}
	\Pi_{\mu\nu}
	=
	-\eta\, \sigma_{\mu\nu}
	+\eta \tau_{\pi}\, \Sigma^{(0)}_{\mu\nu}
	+\lambda_1\, \Sigma^{(1)}_{\mu\nu}
	+\lambda_2\, \Sigma^{(2)}_{\mu\nu}
	+\lambda_3\, \Sigma^{(3)}_{\mu\nu}
	+\lambda_4\, \Sigma^{(4)}_{\mu\nu}
	+\lambda_5\, \Sigma^{(5)}_{\mu\nu}
	+\tilde{\lambda}_1\, \widetilde{\Sigma}^{(1)}_{\mu\nu}
	+\tilde{\lambda}_2\, \widetilde{\Sigma}^{(2)}_{\mu\nu}\ ,
\end{equation}
where
\begin{align}
\notag
    	\Sigma^{(0)}_{\alpha\nu} &= {}_{\langle}u^{\lambda}\partial_{\lambda} \sigma_{\alpha\nu\rangle}+\frac{1}{3}\sigma_{\alpha\nu}\partial_{\lambda}u^{\lambda} \ , \\
\notag
	\Sigma^{(1)}_{\alpha\nu} &=\sigma_{\langle \alpha \lambda}\sigma^{\lambda}_{\phantom{\lambda}\nu\rangle}\ ,\quad
    	\Sigma^{(2)}_{\alpha\nu} =\sigma_{\langle \alpha \lambda}\omega^{\lambda}_{\phantom{\lambda}\nu\rangle}\ ,\quad
    	\Sigma^{(3)}_{\alpha\nu} =\omega_{\langle \alpha \lambda}\omega^{\lambda}_{\phantom{\lambda}\nu\rangle}\ ,\quad
    	\Sigma^{(4)}_{\alpha\nu} =\partial_{\langle \alpha}\frac{\mu}{T} \partial_{\nu\rangle}\frac{\mu}{T}\ ,\\
\notag
        \Sigma^{(5)}_{\alpha\nu} &= \partial_{\langle \alpha} \partial_{\nu \rangle}\frac{\mu}{T}+2 u^\rho \partial_\rho u_{\langle \alpha} \partial_{\nu \rangle}\frac{\mu}{T} -\frac23 \partial_\beta u^\beta u_{\langle \alpha} \partial_{\nu \rangle} \frac{\mu}{T}\ ,\\
	\widetilde{\Sigma}^{(1)}_{\alpha\nu} &= \partial_{\langle \alpha}\frac{\mu}{T}\ell_{\nu\rangle}\ ,\quad
	\widetilde{\Sigma}^{(2)}_{\alpha\nu} = \ell_{\langle \alpha} u^{\gamma}\partial_{\gamma}u_{\nu\rangle}+\frac{1}{2}\partial_{\langle \alpha}\ell_{\nu\rangle}\ .
\label{E:Secondorderterms}
\end{align}

In the rest of this work, we use the AdS/CFT correspondence to compute the various transport coefficients associated with the energy momentum tensor and $R$-charge current of the $\mathcal{N}=4$ theory.\footnote{
In practice our results can be generalized to any CFT whose dual can be truncated to Einstein-Maxwell theory on AdS${}_5$ with a Chern-Simons term. The only difference between the transport coefficients of that CFT and the corresponding ones in the $\mathcal{N}=4$ theory is an overall multiplicative factor of order unity associated with the volume of the compact manifold.}
Sections \ref{S:RNBH} and \ref{S:Main} describe this computation in detail and the results are summarized below.

The energy density and charge density are given by
\begin{subequations}
\label{E:Transport0}
\begin{align}
\label{E:epsilon}
	\epsilon & = \frac{3 N^2}{8 \pi^2 b^4}\\
\label{E:rho}
	\rho & = \frac{\mu r_+^2 N^2}{4\pi^2},
\end{align}
\end{subequations}
where $N$ is the rank of the gauge group
and
\begin{subequations}
\label{E:UsefulIDs}
\begin{align}
        r_+ & = \frac{\pi T}{2} \left(1 + \sqrt{1+\frac23 \frac{\mu^2}{\pi^2 T^2}}\right)\ , \\
	b^{-4} & = \frac{\pi^4 T^4}{2^4}\left(\sqrt{1+\frac23 \frac{\mu^2}{\pi^2 T^2}}+1\right)^3\left(3\sqrt{1+\frac23 \frac{\mu^2}{\pi^2 T^2}}-1\right)\ .
\end{align}
Later we will also need
\begin{equation}
\label{E:rminus}
	r_-^2 = \frac{1}{2}r_+^2
		\left(-1+\sqrt{9-\frac{8}{\frac{1}{2}\left(1+\sqrt{1+\frac{2\mu^2}{3\pi^2 T^2}}\right)}}\right).
\end{equation}
\end{subequations}

The first order transport coefficients are given by
\begin{subequations}
\label{E:Transport1}
\begin{align}
\label{E:eta}
	\frac{\eta}{s} &= \frac{1}{4 \pi}\\
\label{E:kappa}
	\frac{\kappa}{\chi} &= \frac{1}{2} r_+^7 T b^8\\
\label{E:omega}
	\frac{\Omega}{\chi} & = \frac{\mu^2 r_+^4 b^4}{2\sqrt{3}\pi^2 T^2}\,,
\end{align}
\end{subequations}
where $s$ is the entropy density and $\chi$ is the susceptibility,
\begin{equation}
	s =  \frac{1}{3}\frac{\partial\epsilon}{\partial T} = \frac{N^2 r_+^3}{2 \pi},\quad
	\chi = \frac{\partial \rho}{\partial \mu}\Big|_{\mu=0} =  \frac{1}{4} N^2 T^2.
\end{equation}

The thirteen second order transport coefficients are
\begin{subequations}
\label{E:Transport2}
\begin{align}
\label{E:taupi}
	\frac{\eta \tau_{\pi}}{c} & =
		\frac{1+\frac{\mu ^2}{6 r_+^2}}{36 \pi ^2}+\frac{1+\frac{\mu ^2}{6 r_+^2}}{72 \pi ^2  r_+^2  b^4 \left(2
   r_-^2+r_+^2\right)}\ln \left(\frac{r_+^2-r_-^2}{r_-^2+2 r_+^2}\right) \\
\label{E:lambda1}
	\frac{\lambda_1}{c} & = \frac{1}{72 \pi ^2}\left(1+\frac{\mu ^2}{6 r_+^2}\right)\\
\label{E:lambda2}
	\frac{\lambda_2}{c} & = 2 \left(\frac{\eta\tau_{\pi}}{c} -\frac{\left(1+\frac{\mu ^2}{6 r_+^2}\right)}{36 \pi ^2}\right)\\
\label{E:lambda3}
	\frac{\lambda_3}{c} & = -\frac{\mu ^2 b^4 r_+^2}{27 \pi ^2}\left(1+\frac{\mu ^2}{6 r_+^2}\right)\\
\label{E:lambda4}
	\frac{\lambda_4}{c} & = \frac{-1+3 \ln (2)}{216 \pi ^4}\left(1+\mathcal{O}\left(\frac{\mu}{T}\right)\right) \\
\label{E:lambda5}
    \frac{\lambda_5}{c} & = -\frac{ \mu b^8 r_+^4 T^3 }{216}\\
\label{E:tlambda1}
	\frac{\tilde{\lambda}_1}{c} & = 0\\
\label{E:tlambda2}
	\frac{\tilde{\lambda}_2}{c} & = \frac{r_+ \mu ^3  b^4}{54 \sqrt{3} \pi ^2}\left(1+\frac{\mu ^2}{6 r_+^2}\right)
\end{align}
and
\begin{align}
\label{E:xi1}
	\frac{\xi_1}{c} & = \frac{\ln(2)}{72 \pi^4 T}\left(1+\mathcal{O}\left(\frac{\mu}{T}\right)\right)\\
\label{E:xi2}
	\frac{\xi_2}{c} & = \frac{T^3 b^{12}}{144}\left(
		r_-^8+2 r_-^6 r_+^2-r_-^4 r_+^4-2 r_-^2r_+^6+4r_+^8\right)\\
\label{E:xi3}
	\frac{\xi_3}{c} & = \frac{\mu b^7 r_+^2}{48 \pi^2}\left(1+\frac{\mu ^2}{6 r_+^2}\right)\\
\label{E:xi4}
	\frac{\xi_4}{c} & = \frac{\mu^3 r_+^4 b^9}{108\pi^2}\left(1+\frac{\mu^2}{6 r_+^2}\right)\\
\label{E:xit1}
\frac{\tilde{\xi}_1}{c} &= -\frac{\mu ^2 r_+^5 b^8}{72 \sqrt{3}\pi^2}\left(1+\frac{\mu^2}{6 r_+^2}\right)\ ,
\end{align}
\end{subequations}
where we defined
\begin{equation}
	c=\frac{\partial^2 \epsilon}{\partial T^2} = \frac{9 N^2 r_+^2}{2 \left(1+\frac{\mu ^2}{6 r_+^2}\right)}\ .
\end{equation}

Strictly speaking, all our expressions for the transport coefficients are valid when $\mu/T$ is not too large. In terms of the gravity dual from which these coefficients were obtained, this implies that the black hole is not close to extremality.
For brevity, we have omitted the somewhat long expressions for $\lambda_4$ and $\xi_1$ and included only their leading contribution when expanded in a small $\mu/T$ expansion. The interested reader is referred to appendix \ref{A:Long} for the full expressions. A discussion of these results can be found in section \ref{S:Discussion}.
We mention here that the $\mu \to 0$ limit of $\tau_{\pi},\,\lambda_1,\,\lambda_2$ and $\lambda_3$ coincides with the computation of \cite{Baier:2007ix,Bhattacharyya:2007jc,Natsuume:2007ty} for the $\mu=0$ case. In \cite{Natsuume:2007ty} the dispersion relations for the current were computed via the Kubo formula and compared to the Israel-Stewart theory.


\section{Setup}
\label{S:RNBH}
In the previous section we have explained how one can define the hydrodynamic transport coefficients of a conformal theory from the form of the energy momentum tensor and current. In what follows we explain how these transport coefficients can be computed from the bulk dual of the gauge theory.

Our starting point is the five dimensional action of Einstein-Maxwell
theory
\begin{equation}
\label{E:action}
	S = -\frac{1}{16 \pi G_5}\int \left[ \sqrt{-g}\left(R + 12
   - \frac{1}{4} F^2\right) - \frac{1}{12\sqrt{3}}\epsilon^{MNOPQ}A_{M}
   F_{NO}F_{PQ}\right]d^5x\ .
\end{equation}
The metric
\begin{equation}
\label{E:StaticRNBH}
	ds^2 = -r^2 f(r) u_{\mu}u_{\nu}dx^{\mu}dx^{\nu}
	       +r^2 P_{\mu\nu} dx^{\mu}dx^{\nu}
	       - 2 u_{\mu}dx^{\mu}dr
\end{equation}
with $u^{\mu}$ a fixed four vector satisfying $u^{\mu}u_{\mu}=-1$,
\begin{equation}
\label{E:deff}
        f(r)=1+\frac{Q^2}{r^6}-\frac{1}{b^4 r^4}
\end{equation}
and $P_{\mu\nu}$ as in \eqref{E:defProjection},
together with the gauge field
\begin{equation}
\label{E:StaticRNGF}
	A_r = 0\ , \quad A_\mu = -\frac{\sqrt{3}Q}{r^2} u_\mu
\end{equation}
are solutions to the Einstein-Maxwell equations
\begin{eqnarray}
\label{E:EinsteinMaxwell}
	R_{MN}+4 g_{MN}&=&\frac12 F_{MK} F_N\, ^K - \frac{1}{12} g_{MN} F^2\ ,
        \nonumber \\
        \partial_N(\sqrt{-g} F^{NM}) &=& \frac{1}{4\sqrt{3}}\epsilon^{MNOPQ}F_{NO}F_{PQ}\
\end{eqnarray}
derived from \eqref{E:action}.\footnote{Note that we do not require the gauge field to vanish at the future horizon and therefore it is likely that it diverges at the past horizon. In fact, it is likely that the whole perturbative solution diverges at the past horizon because generic solutions of viscous fluid dynamics are not expected to be regular in the infinite past. We thank A. Karch, D. Son, and especially R. Loganayagam for clarifying this point.} In what follows, Greek indices run over the boundary coordinates $\mu = 0,\ldots,3$ while Roman indices run over the bulk coordinates $N=0,\ldots,4$.

The metric \eqref{E:StaticRNBH} is nothing but a boosted version of the charged black brane solution expressed in the Eddington-Finkelstein coordinate system. We would like to extend the solution \eqref{E:StaticRNBH} and \eqref{E:StaticRNGF} by allowing $u^{\mu}$, $Q$ and $b$ to vary slowly with the space-time coordinates. It is a simple exercise to check that \eqref{E:StaticRNBH} and \eqref{E:StaticRNGF} are no longer solutions to the Einstein-Maxwell equations \eqref{E:EinsteinMaxwell} once the boost parameters, charge, and mass of the black hole are allowed to vary. Thus, we need to correct the metric \eqref{E:StaticRNBH} and \eqref{E:StaticRNGF} to take into account the change in $u^{\mu}$, $b$ and $Q$. We will do this order by order in a derivative expansion.
To set the stage for our perturbative expansion we decompose our metric and gauge field into scalars, vectors and tensors with respect to the local velocity field of the fluid,
\begin{multline}
\label{E:Correctedmetric}
	ds^2 = r^2 k(r) u_{\mu}u_{\nu}dx^{\mu}dx^{\nu}
	       +r^2 h(r) P_{\mu\nu} dx^{\mu}dx^{\nu}
	       +r^2 \pi_{\mu\nu}(r) dx^{\mu}dx^{\nu} \\
           +r^2 j_{\sigma}(r)\left(P^{\sigma}_{\mu}u_{\nu}+P^{\sigma}_{\nu}u_{\mu}\right)dx^{\mu}dx^{\nu}
           -2 S(r) u_{\mu}dx^{\mu}dr\\
	   \equiv
	   r^2 g_{\mu\nu}dx^{\mu}dx^{\nu}-2 S(r)u_{\mu}dx^{\mu}dr\ ,
\end{multline}
and
\begin{equation}
\label{E:Correctedvector}
	A_r = 0\ , \quad A_\mu = a_\nu(r) P^{\nu}_{\mu}  + c(r) u_\mu\ .
\end{equation}
We point out that the various functions $k(r)$, $h(r)$, etc. are not only functions of the radial coordinate $r$ but could also depend, in principle, on the charge, $Q(x^{\alpha})$, the mass parameter $b(x^{\alpha})$ and the velocity field $u^{\mu}(x^{\alpha})$, or on their derivatives. So $k(r)$, $h(r)$, $\ldots$ implicitly depend on the transverse coordinates.
We have chosen an axial gauge for the gauge field, $A_{r}=0$, and set $g_{\mu r}\propto u_{\mu}$ with $g_{rr}=0$. There is one extra gauge degree of freedom which we will fix shortly. Our goal is to compute the functions $k,\,h,\,\pi_{\mu\nu},\,j_{\alpha},\,S,\,a_{\nu}$ and $c$ order by order in a derivative expansion of $Q(x^{\alpha})$, $b(x^{\alpha})$ and $u^{\mu}(x^{\alpha})$. Using a superscript $(n)$ to denote the $n$'th order contribution to such an expansion, we can rewrite \eqref{E:StaticRNBH} and \eqref{E:StaticRNGF} as
\begin{align}
\label{E:zeroorder}
    k^{(0)}(r) &= -f(r)\ ,&
    S^{(0)}(r) &= 1\ ,&
    h^{(0)}(r) &= 1\ ,\nonumber \\
    j_{\mu}^{(0)}(r) &= 0\ ,&
    \pi^{(0)}_{\mu\nu}(r) &= 0\ , \\
    c^{(0)}(r) &= -\frac{\sqrt{3}Q}{r^2}\ ,&
    a_{\mu}^{(0)}(r) &= 0\ . \nonumber
\end{align}

A computation of the various functions $k(r),\,h(r),\,\ldots$ is carried out in section \ref{S:Main}. Once these are obtained, we can compute the energy momentum tensor $\langle T_{\mu\nu} \rangle$ and $R$-charged current $\langle J_{\mu} \rangle$ of the boundary theory using the standard AdS/CFT dictionary \cite{Gubser:1998bc,Witten:1998qj,Balasubramanian:1999re,deHaro:2000xn,Bianchi:2001de,Bianchi:2001kw} adopted to Eddington-Finkelstein coordinates \cite{Haack:2008cp}. In the Landau gauge this reads
\begin{align}
\notag
	16\pi G_5\langle T_{\mu\nu}\rangle  =& k^{(\overline{4})} \left(4 u_{\mu}u_{\nu}+\eta_{\mu\nu}\right)
		+ 4 \pi_{\mu\nu}^{(\overline{4})}\\
	\langle J^{\mu} \rangle
	=&
	\frac{1}{\sqrt{-g^{(\overline{0})}}}\frac{\delta}{\delta A_{\mu}^{(\overline{0})}}
	S_{\rm ren}[A_{\mu}^{(\overline{0})},g_{\mu\nu}^{(\overline{0})}]
	=
	-\frac{1}{8\pi G_5} \eta^{\rho \mu} a^{(\overline{2})}_\rho,
\label{E:BtB}
\end{align}
where a barred superscript $(\overline{n})$ indicates the $n$'th term in a large $r$ (near boundary) expansion of the appropriate expression. Working in the Landau frame also requires that
\begin{equation}
\label{E:Landaugauge}
 	j^{(\overline{4},n)}_\mu =0\, \quad k^{(\overline{4},n)}=0\, \quad
	c^{(\overline{2},n)} =0
\end{equation}
for $n \geq 1$. If \eqref{E:Landaugauge} is not satisfied this would correspond to a small shift in the local velocity fields, the energy and charge densities.

The chemical potential $\mu$ of the boundary theory is given by the difference between the value of the temporal component of the
gauge field at the horizon and its value at the boundary
(of the unboosted black hole solution).
The temperature $T$ of the boundary theory can be
obtained from the Hawking temperature,
\begin{equation}
\label{E:Tandmu}
	\mu = A_t(r_+)-A_t(\infty) = \frac{\sqrt{3}Q}{r_+^2},\quad
	 T=\frac{r_+}{2\pi}\left(2-\left(\frac{r_-}{r_+}\right)^2-\left(\frac{r_-}{r_+}\right)^4 \right)\ .
\end{equation}
We have defined $r_+$ to be the larger of the two positive roots of $f(r)$ and $r_-$ the smaller of its two positive roots (the other four roots of $f(r)$
are given by $-r_+$, $-r_-$ and $\pm i \sqrt{r_+^2+r_-^2}$).
By manipulating \eqref{E:Tandmu} and \eqref{E:deff} one can obtain \eqref{E:UsefulIDs}.
Note that an extremal black hole is obtained when $r_+=r_-$. In the
boundary theory this corresponds to the limit $T \to 0$, while keeping
$\mu \neq 0$, as can be seen from \eqref{E:rminus}.
In \cite{Bhattacharyya:2007vs} it was shown that in this limit, even though the temperature vanishes, the mean free path is not necessarily vanishing and one might expect a hydrodynamic description of the theory in this regime. However, following \cite{Gubser:2000ec}, it was also argued that this regime of the theory is likely to be unstable. We will also see shortly that, from a bulk point of view, our
perturbative analysis breaks down when the black hole is close to extremality.
Hence, in what follows we will assume that $\mu/T \ll 1$.

It is now straightforward to derive the energy density $\epsilon$ in \eqref{E:epsilon} and the charge density $\rho$ in \eqref{E:rho} by inserting \eqref{E:zeroorder} into \eqref{E:BtB}
and using
\begin{equation}
	16\pi G_5 = \frac{8 {\rm Vol}_5}{\pi N^2}\ ,
\end{equation}
where ${\rm Vol}_5 = \pi^3$ for the $\mathcal{N}=4$ theory.


\section{The derivative expansion}
\label{S:Main}

If $u^{\mu}$, $Q$ and $b$ are constants then \eqref{E:zeroorder} is a solution to \eqref{E:EinsteinMaxwell}. As emphasized earlier, if we allow the fields $u^{\mu}$, $b$ and $Q$ to vary with the space-time coordinates then \eqref{E:zeroorder} is no longer a solution to the equations of motion. However, if we allow $u^{\mu}$, $b$ and $Q$ to vary slowly in the transverse coordinates then we can construct a solution perturbatively. At zero order we have the solution \eqref{E:zeroorder}. At first order, we look for a correction to \eqref{E:zeroorder}, expressed in terms of functions $k^{(1)}(r),\,h^{(1)}(r),\,\pi_{\mu\nu}^{(1)}(r),\,j_{\alpha}^{(1)}(r),\,S^{(1)}(r),\,a_{\nu}^{(1)}(r)$ and $c^{(1)}(r)$ which depend on one derivative of the hydrodynamic fields. That is, we insert \eqref{E:Correctedmetric} and \eqref{E:Correctedvector}, with $k(r)=k^{(0)}(r)+k^{(1)}(r)$, $h(r)=h^{(0)}(r)+h^{(1)}(r)$, etc., into the Einstein-Maxwell equations \eqref{E:EinsteinMaxwell}, omitting all terms which contain two or more derivatives of the charge, temperature or velocity fields. These equations will of course be linear in $k^{(1)}(r),\,h^{(1)}(r)$, etc. If these equations can be solved then the solution will give us the metric and gauge field of a charged AdS${}_5$ black hole, where the charge, mass and boost parameters slowly vary in the transverse coordinates, valid to first order in gradients of these parameters. With the first order solutions at hand, this procedure may be repeated to obtain $k^{(2)}(r)$, $h^{(2)}(r)$, etc.---a solution to the Einstein-Maxwell equations involving two derivatives of the hydrodynamic fields. Up to some caveats which we discuss below, one may, in principle, carry out this algorithm to an arbitrary order in the derivative expansion.

It is straightforward, but tedious to compute the order $n$ Einstein-Maxwell equations. A method developed in \cite{Bhattacharyya:2007jc} which simplifies this task is to consider the equations of motion in the neighborhood of a point $x^{\mu}_0$ but at arbitrary radial coordinate $r$. The hydrodynamic fields are expanded in a Taylor series around $x^{\mu}_0$ up to order $n$. Thus, no information is lost regarding an order $n$ derivative expansion. Once a solution is obtained around $x^{\mu}_0$ it can be uniquely extended to the entire manifold. The interested reader is referred to \cite{Bhattacharyya:2007jc} for an extended discussion of this method. As in \cite{Bhattacharyya:2007jc} we choose $x^{\mu}_0 = 0$. At this point we can also choose $u^{\mu} = (1,0,0,0)$, $b=b_0$ and $Q=Q_0$.

After implementing this technique, we find that in the neighborhood of $x_0^{\mu}$ the Einstein-Maxwell equations \eqref{E:EinsteinMaxwell} take the form
\begin{subequations}
\label{E:EinsteinEOM}
\begin{align}
\label{E:EOMPI}
	\partial_r\left(r^5 f(r) \partial_r \pi^{(n)}_{ij}\right) &= \mathbf{P}^{(n)}_{ij}(r)\\
\label{E:EOMJ}
	\partial_r \left(r^5\partial_r j^{(n)}_i(r)+2\sqrt{3} Q_0 a^{(n)}_i(r) \right)
	&=\mathbf{J}^{(n)}_{i}(r)\\
\label{E:EOMS}
	3 \partial_r S^{(n)}(r) - \frac{3}{2} r^{-1}\partial_r \left(r^2 \partial_r h^{(n)}(r)\right)&=\mathbf{S}^{(n)}(r)\\
\label{E:EOMK}
	\partial_r\left(r^4 k^{(n)}(r)\right)+8 r^3 S^{(n)}(r)+
	b_0^{-4}\left(1-3 r^4 b_0^4\right)\partial_r h^{(n)}(r) - \frac{2}{\sqrt{3}} Q_0 \partial_r c^{(n)}
	&= \mathbf{K}^{(n)}(r)\,,
\end{align}
\end{subequations}
and
\begin{subequations}
\label{E:MaxwellEOM}
\begin{align}
\label{E:EOMC}
        \partial_r \left( r^3 \partial_r c^{(n)} \right) - 2\sqrt{3} Q_0 \partial_r S^{(n)} + 3\sqrt{3} Q_0 \partial_r h^{(n)} &= \mathbf{C}^{(n)}(r)\\
\label{E:EOMA}
        \partial_r \left(r^3 f(r) \partial_r a^{(n)}_i(r)+2\sqrt{3}L Q_0 j^{(n)}_i(r) \right)  &= \mathbf{A}_i^{(n)}(r)
\end{align}
\end{subequations}
at order $n$ in a derivative expansion.
In addition, there are four constraint equations which restrict the allowed values of $Q(x^{\alpha})$, $b(x^{\alpha})$ and $u^{\mu}(x^{\alpha})$ and reduce to the conservation equations \eqref{E:Conservation} when expanded to order $n$.

While the ``kinetic'' terms for the unknown fields $k^{(n)}(r)$, $h^{(n)}(r)$, etc are identical for all $n$, the source terms, on the right hand side of \eqref{E:EinsteinEOM} and \eqref{E:MaxwellEOM} must be determined at every order.
We obtain the explicit form of the $n=1$ and $n=2$ sources in the next section. Once the sources are known, it is simply a manner of integrating the equations of motion \eqref{E:EinsteinEOM} and \eqref{E:MaxwellEOM} to obtain a solution at order $n$.

For the  tensor modes $\pi^{(n)}_{\mu\nu}$ we find
\begin{equation}
\label{E:piIntegral}
    \pi^{(n)}_{\mu \nu}(r) = -\int_{r}^{\infty}\frac{\int_{r_+}^x \mathbf{P}^{(n)}_{\mu \nu}(x^{\prime})dx^\prime}{x^5f(x)} dx\ ,
\end{equation}
where $\pi^{(n)}_{\mu \nu}$ and $\mathbf{P}^{(n)}_{\mu \nu}$ reduce to
$\pi^{(n)}_{ij}$ and $\mathbf{P}^{(n)}_{ij}$ of \eqref{E:EOMPI} when
expanded around $x_0^\mu=0$.
The boundary conditions we have imposed are that the boundary metric remain flat, i.e., the bulk metric is not deformed near the boundary, and that all singularities are veiled behind the outer horizon roughly located at $r=r_+$. The upper limit of the outer integral in \eqref{E:piIntegral} ensures that the former boundary condition is satisfied. The lower limit of the inner integral in \eqref{E:piIntegral} ensures that the outer integrand remains finite at $r = r_+$ where $f(r_+)=0$. As is standard for charged black holes, once the solution \eqref{E:zeroorder} is perturbed, the inner horizon, located at $r\sim r_-$, becomes singular (see for example \cite{Townsend:1997ku}). As long as this singularity is located behind the outer horizon, we should not worry about this. However, since the outer horizon is no longer located precisely at $r = r_+$, and since we do not want the fluctuations of the horizon to reveal the singularity at $r=r_-$, we require that $r_+ \gg r_-$. 
More details about the geometry of the perturbed horizon (in the uncharged case) can be 
found in \cite{Bhattacharyya:2008xc} 

In order to determine the energy momentum tensor of the boundary theory, we do not need $\pi_{\mu\nu}(r)$ but only its fourth order term in a near boundary expansion $\pi^{(\overline{4})}_{\mu\nu}$, c.f., \eqref{E:BtB}. From \eqref{E:piIntegral} one finds
\begin{equation}
\label{E:pitrick}
	4 \pi^{(\overline{4})}_{\mu\nu}
	=
	\lim_{r\to\infty}
	\left[\sum_{m=0}^2 (-1)^m \frac{r^{m+1} \partial^m_r \mathbf{P}_{\mu\nu}(r)}{(m+1)!}-\int_{r_+}^r \mathbf{P}_{\mu\nu}(x)dx\right].
\end{equation}
Thus, if it is only the order $n$ boundary theory energy momentum tensor we are looking for, $\pi_{\mu\nu}^{(\overline{4},n)}$, we are excused from doing the double integral in \eqref{E:piIntegral}.

Before considering the integral solutions for the scalar modes, $k^{(n)}(r)$, $S^{(n)}(r)$, $c^{(n)}(r)$ and $h^{(n)}(r)$, we recall that we have not completely fixed our gauge for the metric. As in \cite{Haack:2008cp}, we choose the gauge $h(r)=1$ since it allows us to easily decouple $k^{(n)}(r)$, $S^{(n)}(r)$ and $c^{(n)}(r)$. Other possible gauges are $S(r)=1$ and $S(r) = -3/2 h(r)$, which were used in \cite{Bhattacharyya:2007jc,Bhattacharyya:2008xc}.
After choosing the gauge $h(r)=1$, we find
\begin{align}
\notag
    S^{(n)}(r) &=-\frac{1}{3}\int_r^{\infty} \mathbf{S}^{(n)}(x) dx \\
\notag
    c^{(n)}(r) &=-\int_{r}^{\infty}x^{-3}\int_{r^+}^x \left[ \mathbf{C}^{(n)}(x^{\prime})+\frac{2}{\sqrt{3}}Q \mathbf{S}^{(n)}(x^{\prime}) \right] dx^{\prime} dx + c_0 r^{-2}\\
    k^{(n)}(r) &=r^{-4} \int_{r^+}^{r} \left[ \mathbf{K}^{(n)}(x)
    -8 r^3 S^{(n)}(x) + \frac{2}{\sqrt{3}} Q \partial_r c^{(n)}(x) \right] dx + C_0 r^{-4}\ ,
\label{E:ScalarIntegral}
\end{align}
where we required again that the boundary metric is flat and that there are no singularities for $r\ge r_+$.
The sources $\mathbf{S}$, $\mathbf{K}$ and $\mathbf{C}$ reduce to the ones in \eqref{E:EOMS}, \eqref{E:EOMK} and \eqref{E:EOMC} when expanded around $x_0^{\mu}$.
The extra integration constants $C_0$ and $c_0$ are fixed by the Landau gauge \eqref{E:Landaugauge}. As we discussed in sections \ref{S:Conformalfd} and \ref{S:RNBH}, the charge density and energy density were defined relative to the rest frame of a fluid element. This definition was a result of our choice of frame. In the gravity dual this choice manifests itself as a choice of adding an extra zero momentum quasi-normal mode to $c^{(n)}(r)$ and $k^{(n)}(r)$. Choosing the Landau frame implies choosing $c^{(\overline{2},n)}=0$ and $k^{(\overline{4},n)}=0$ in the bulk theory for $n\ge 1$. Thus, we choose values for $c_0$ and $C_0$ such that the second term in a near boundary expansion of $c^{(n)}(r)$ and the fourth order term in a near boundary expansion of $k^{(n)}(r)$ vanish for $n \geq 1$. We refer the reader to \cite{Bhattacharyya:2007jc,Haack:2008cp} for details.

For the vector equations we find
\begin{equation}
\label{E:Vmodes}
	\begin{pmatrix}
	 	a^{(n)}_{\nu}(r) \\ j^{(n)}_{\nu}(r)
	\end{pmatrix}
		=
	\left(-H(r) \int_r^{\infty}
	 H^{-1}(x)
	\begin{pmatrix}
	 	\frac{\int^x_{r_+} \mathbf{A}_{\nu}(x^{\prime})dx'}{x^3 f(x)} \\
		x^{-5}\int^x_{r_+} \mathbf{J}_{\nu}(x^{\prime})dx'
	\end{pmatrix}
	dx\right)
	+
	C_{\nu} H_1(r) + D_{\nu} H_2(r)
\end{equation}
after extending the solution to the entire manifold.
The $\coefone_{\nu}$'s are chosen so that the $r^{-4}$ terms in a near
boundary expansion of $j(r)_\nu$ vanish---again a result of our working in the Landau frame. The
coefficients $\coeftwo_{\nu}$ are chosen so that $a^{(n)}_{\mu}(r)$ and $j^{(n)}_{\nu}(r)$ will be
finite at the horizon. The columns of the matrix $H(r)$ are given by the
solutions to the homogeneous version of the vector equations, $H_1(r)$ and $H_2(r)$.
Its explicit form and a detailed discussion of the solution \eqref{E:Vmodes} and its derivation can be found in appendix \ref{A:Vector}. As was the case for the tensor modes, in order to compute the order $n$ contribution to the $R$-charge current we do not need the full solution $a^{(n)}_\nu(r)$ but only its $r^2$ coefficient in a near boundary expansion. From \eqref{E:Vmodes} we find
\begin{equation}
\label{E:A2value}
	a^{(\overline{2})}_{\nu} = \lim_{r\to \infty}
	\left[
	    \frac{1}{2}\left(r \mathbf{A}_{\nu}(r)-\int_{r^+}^r \mathbf{A}_{\nu}(x)dx\right)
	\right]
        -\sqrt{3} b^4 Q\, \coefone_{\nu}
        -\frac{\sqrt{3}}{4} b^4 Q\, \coeftwo_{\nu}
\end{equation}
where $\coefone_{\nu}$ and $\coeftwo_{\nu}$ are given by
\begin{align}
	4 \coefone_{\nu} &=-\lim_{r\to\infty}
    \Bigg[ \sum_{m=0}^2 \frac{(-1)^{m}r^{m+1}}{(m+1)!}\partial_r^m \mathbf{J}_{\nu}(r)-\int_{r_+}^r \mathbf{J}_{\nu}(x)dx \Bigg]\\
\intertext{and}
\label{E:alphasimple}
	\coeftwo_{\nu} &= -
	\sqrt{3} Q\int_{r^+}^{\infty} \frac{\mathbf{A}_{\nu}(x)}{x^2}dx
	-
	\frac{1}{b^4}\int_{r^+}^{\infty} \frac{\mathbf{J}_{\nu}(x)}{x^4}dx
	+Q^2 \int_{r^+}^{\infty}\frac{\mathbf{J}_{\nu}(x)}{x^6}dx\ .
\end{align}

One caveat in the previous analysis is that we have assumed that a solution exists,
i.e., that all the integrals are well defined. However, if the sources are too
divergent near the boundary we would find that the outer integrals in
\eqref{E:piIntegral}, \eqref{E:ScalarIntegral} and \eqref{E:Vmodes} do not exist. This would imply that one could not
impose the boundary condition that the boundary metric remains flat.
By inspection,
one can check that the order $n$ sources for the metric and gauge field will
not deform the boundary theory as long as
\begin{align}
\notag
 	\mathbf{P}^{(n)}_{\mu\nu}(r)=\mathcal{O}(r^2),\quad
	\mathbf{S}^{(n)}(r)=\mathcal{O}(r^{-2}),\quad
	\mathbf{K}^{(n)}(r) =\mathcal{O}(r^2)\\
\label{E:Nodeform}
	\mathbf{J}_{\nu}^{(n)}(r) = \mathcal{O}(r^2),\quad
	\mathbf{C}^{(n)}(r) = \mathcal{O}(r^0),\quad
	\mathbf{A}_{\nu}^{(n)}(r) = \mathcal{O}(r^0).
\end{align}
We have checked that this is the case up to order $n=2$.


\subsection{First order expansion}
It remains to evaluate the various sources, $\mathbf{P}_{\mu\nu},\,\mathbf{S},\,\mathbf{K},\,\mathbf{J}_{\nu},\,\mathbf{C}$, and $\mathbf{A}_\nu$, order by order in a derivative expansion.
By direct computation, the first order sources are given by
\begin{subequations}
\label{E:SourcesO1}
\begin{align}
    \mathbf{P}^{(1)}_{\mu\nu}(r) & = -3 r^2 \sigma_{\mu\nu}\\
    \mathbf{S}^{(1)}(r)&=0\\
    \mathbf{K}^{(1)}(r)&=2 r^2 \partial_{\mu}u^{\mu}\\
\label{E:j1source}
    \mathbf{J}^{(1)}_{\nu}(r)&=3 r^2 u^{\mu}\partial_{\mu}u_{\nu}\\
    \mathbf{C}^{(1)}(r)&=0\\
\label{E:Asource}
    \mathbf{A}^{(1)}_{\nu}(r)&=-\frac{\sqrt{3}}{r^2}\left(P_{\nu}^{\alpha}\partial_{\alpha}Q+Q u^{\mu}\partial_{\mu} u_{\nu}\right)
		+\frac{4\sqrt{3}Q^2}{r^5}\ell_{\nu}\ ,
\end{align}
\end{subequations}
when restricted to the neighborhood of $x^{\mu}_0$. After verifying that \eqref{E:Nodeform} holds,
we can use \eqref{E:A2value} to compute the $r^{-2}$ coefficient of $a^{(1)}_\nu(r)$. Before doing so, we make a few remarks on the index structure of the vector modes.
Under the conservation law \eqref{E:Conservation}, the source terms for the vectors can be rewritten as
\begin{align}
\notag
	\mathbf{J}^{(1)}_{\nu}(r) =& \mathbf{J}^{Du}(r)\, u^{\mu}\partial_{\mu}u_{\nu}\\
\label{E:Vsources}
	\mathbf{A}^{(1)}_{\nu}(r) =& \mathbf{A}^{Du}(r)\, u^{\mu}\partial_{\mu}u_{\nu}
				+\mathbf{A}^{\kappa}(r)\, P^{\alpha}_{\,\nu}\partial_{\alpha}\frac{\mu}{T}
				+\mathbf{A}^{\Omega}(r)\,\ell_{\nu}\ ,
\end{align}
where
\begin{align}
	\mathbf{A}^{Du}(r) &= \frac{2\sqrt{3}Q}{r^2} &
    \mathbf{J}^{Du}(r) &= 3 r^2 \\
    \mathbf{A}^{\Omega}(r) &= \frac{4\sqrt{3} Q^2}{r^5} &
	\mathbf{A}^{\kappa}(r) &=-\frac{\pi^2 T^3 r_+^4 b^4}{r^2
        \left(1+\frac{\mu^2}{6 r_+^2} \right)}.
\end{align}
Since the differential equations \eqref{E:EinsteinEOM} and \eqref{E:MaxwellEOM} are linear differential equations in the radial variable $r$, the index structure of the sources in the transverse dimensions will carry through to the gauge field $a^{(1)}_\nu(r)$. Thus, to first order in the derivative expansion, we already see from \eqref{E:Vsources} that $a^{(\overline{2})}_\nu$ may be decomposed into terms proportional to $\ell_{\nu}$, $P^{\alpha}_{\,\nu}\partial_{\alpha}\frac{\mu}{T}$ and $u^{\alpha}\partial_{\alpha}u_{\nu}$. Since $a^{(\overline{2})}_\nu$ is proportional to the boundary current $\langle J_{\mu} \rangle$ via \eqref{E:BtB}, and the boundary current is Weyl invariant, this implies that the non-Weyl invariant term $u^{\alpha}\partial_{\alpha}u_{\nu}$ can not contribute to $a^{(\overline{2})}_\nu$. Indeed, from \eqref{E:A2value} we find
\begin{equation}
\label{E:a2order1}
	a^{(\overline{2})}_{\nu} = -\sqrt{3}Qu_{\nu}+\frac{1}{2}\pi^2 r_+^7 T^3 b^8 \partial_{\nu}\frac{\mu}{T}-\frac{\sqrt{3}}{2}Q^2b^4\ell_{\nu}.
\end{equation}

Using \eqref{E:BtB} we see that the expectation value of the $R$-charge current $\langle J_{\mu} \rangle$ takes the form \eqref{E:GeneralJ} with $\Omega$ and $\kappa$ as in \eqref{E:omega} and \eqref{E:kappa}.
The energy momentum tensor can be evaluated in a similar manner. We compute
$\pi^{(\overline{4})}_{\mu \nu}$ from \eqref{E:pitrick},  and from \eqref{E:BtB} we obtain $\langle T_{\mu\nu} \rangle$. Not surprisingly, it takes the form \eqref{E:GeneralT} with a shear viscosity $\eta$ as in \eqref{E:eta}.

Had we been content with the first derivative corrections to the energy momentum tensor and current, we could have stopped here. Since we will be computing also the second order corrections, we need the full gravity solution to first order in a derivative expansion.
Almost all the sources for the scalar terms in the metric are trivial, and integrating them gives us
\begin{equation}
    S^{(1)}(r)=0\qquad
    c^{(1)}(r)=0\qquad
    k^{(1)}(r)=\frac{2}{3r}\partial_{\mu}u^{\mu}\ .
\end{equation}
For the tensor modes we find
\begin{equation}
    \pi^{(1)}_{\mu\nu}(r)= F(r) \sigma_{\mu\nu}\ ,
\end{equation}
where
\begin{equation}
 	F(r) = \frac{1}{2} b^4 \sum_x \frac{(r_+^3-x^3)\ln(r-x)}{3 b^4 x^4-1}
	  = \frac{1}{r}-\frac{r_+^3}{4r^4}+\mathcal{O}(r^{-5})
\end{equation}
and the sum runs over all six roots of $f(r)$ which was defined in \eqref{E:deff}, i.e.\
$\pm r_+, \pm r_-$ and $\pm i \sqrt{r_+^2+r_-^2}$.
Note that $F(r)$ depends implicitly  on the transverse coordinates.
The expressions for the vector components of the metric are somewhat more involved.
By explicitly carrying out the integral in \eqref{E:Vmodes} we find
\begin{align}
\notag
    j^{(1)}_{\nu}(r)=&-\frac{1}{r} u^{\mu}\partial_{\mu} u_{\nu} -\frac{b^4 Q^3}{2 r^6} \ell_{\nu}+j^{\kappa}(r)\partial_{\nu}\frac{\mu}{T} \\
    a^{(1)}_{\nu}(r)=&
	a^{\kappa}(r)\partial_{\nu}\frac{\mu}{T}
		-\frac{\sqrt{3} Q^2 b^4}{2 r^2} \ell_{\nu}\ ,
\label{E:j1a1}
\end{align}
where $a^{\kappa}(r)$ and $j^{\kappa}(r)$ are given by rather long expressions which we avoid writing out explicitly here and which are given in appendix \ref{A:Long}, equations \eqref{E:akappalong} and \eqref{E:jkappalong}. Their leading order behavior is given by
\begin{align}
\notag
 	a^{\kappa}(r) &= \frac{\pi^2 r_+^7 T^3 b^8}{2 r^2} +\mathcal{O}(r^{-3})\ ,\\
	j^{\kappa}(r) &= \mathcal{O}(r^{-6})\ .
\end{align}


\subsection{Second order}
At second order we find that the scalar sources for the Einstein-Maxwell equations \eqref{E:MaxwellEOM} and \eqref{E:EinsteinEOM} satisfy \eqref{E:Nodeform}
as required. The sources for the tensor modes are given by
\begin{equation}
 	\mathbf{P}^{(2)}_{\mu\nu}=
		\mathbf{P}^{\tau_\pi}\Sigma_{\mu\nu}^{(0)}
		+\mathbf{P}^{\lambda_1}\Sigma_{\mu\nu}^{(1)}
		+\mathbf{P}^{\lambda_2}\Sigma_{\mu\nu}^{(2)}
		+\mathbf{P}^{\lambda_3}\Sigma_{\mu\nu}^{(3)}
		+\mathbf{P}^{\lambda_4}\Sigma_{\mu\nu}^{(4)}
		+\mathbf{P}^{\lambda_5}\Sigma_{\mu\nu}^{(5)}
		+\mathbf{P}^{\tilde{\lambda}_1}\widetilde{\Sigma}_{\mu\nu}^{(1)}
		+\mathbf{P}^{\tilde{\lambda}_2}\widetilde{\Sigma}_{\mu\nu}^{(2)}
\end{equation}
with
\begin{subequations}
\label{E:EMTsources}
\begin{align}
 	\mathbf{P}^{\tau_\pi}(r)&=r-\left(r^3 F(r)\right)^{\prime}-r^3 F^{\prime}(r) \\
	\mathbf{P}^{\lambda_1}(r)&=r-3 r^2 F(r)-(r^3-r_+^3)F^{\prime}(r) \\
	\mathbf{P}^{\lambda_2}(r)&=2\left(r-\left(r^3 F(r)\right)^{\prime}-r^3 F^{\prime}(r)\right)-4 r \\
\label{E:Plambda3}
    \mathbf{P}^{\lambda_3}(r)&= 4r \left(1+\frac{1}{b^4 r^4}+\frac{2 Q^2}{r^6}\right)
        +\frac{4 Q^4 b^4}{r^3}\left(
            -3 b^4+\frac{3}{r^4}-\frac{12 Q^2 b^4}{r^6}\right)\\
\notag
	\mathbf{P}^{\lambda_4}(r)&=-\frac32 T \frac{(\pi T)^2 + r^2 -\tfrac23 \pi T r}{\pi \left(r^2+(\pi T)^2\right)} \\	
\label{E:Plambda4}
	&\phantom{=}+
	\left(\frac{\pi  T^3}{4 r^2}+\frac{3 r^2}{4 \pi ^3 T}\right) \left(\ln\left(\frac{r^2 + (\pi T)^2}{(r + \pi T)^2} \right) - 2 \arctan\left(\frac{r}{\pi T} \right) + \pi \right)+\mathcal{O}\left(\mu\right) \\
    \mathbf{P}^{\lambda_5}(r)&= -2 \left(r^3 j^{\kappa}(r)\right)^{\prime}\\
    \mathbf{P}^{\tilde{\lambda}_1}(r)&=-\frac{\sqrt{3}Q^2 b^8 T^3 \pi^2
        r_+^{4}}{\left(1+\frac{\mu^2}{6 r_+^2}\right)r^4}
		+\frac{\sqrt{3}Q^2b^4}{r^6}
		\left(\frac{12 Q^2}{r}a^{\kappa}(r)-2 r^6 f(r) {a^{\kappa}}^{\prime}(r)\right)\\
	\mathbf{P}^{\tilde{\lambda}_2}(r)&=-\frac{6 Q^3 b^4}{r^4}\ .
\end{align}
\end{subequations}
The coefficients $\lambda_i$ and $\tau_{\pi}$ of the energy momentum tensor can be computed by inserting the corresponding sources of \eqref{E:EMTsources} into \eqref{E:pitrick} and using \eqref{E:BtB}. The results are listed in \eqref{E:Transport2}.
The full expressions for $\mathbf{P}^{\lambda_4}(r)$ can be found in 
\eqref{E:Plambda4long}.

The sources for the vector modes are given by
\begin{align}
	\mathbf{A}^{(2)}_{\nu}=&	
			\mathbf{A}^{\xi_1}\Xi^{(1)}_{\nu}
			+\mathbf{A}^{\xi_2}\Xi^{(2)}_{\nu}
			+\mathbf{A}^{\xi_3} \Xi^{(3)}_{\nu}
			+\mathbf{A}^{\xi_4}\Xi^{(4)}_{\nu}
			+\mathbf{A}^{\tilde{\xi}_1}\widetilde{\Xi}^{(1)}_{\nu} \nonumber \\
	\mathbf{J}^{(2)}_{\nu}=&
			\mathbf{J}^{\xi_1}\Xi^{(1)}_{\nu}
			+\mathbf{J}^{\xi_2}\Xi^{(2)}_{\nu}
			+\mathbf{J}^{\xi_3}\Xi^{(3)}_{\nu}
			+\mathbf{J}^{\xi_4}\Xi^{(4)}_{\nu}
			+\mathbf{J}^{\tilde{\xi}_1}\widetilde{\Xi}^{(1)}_{\nu}\ ,
\end{align}
where
\begin{subequations}
\label{E:Asources}
\begin{align}
	\mathbf{A}^{\xi_1}(r) &= r {a^{\kappa}}^{\prime}(r)-T^3 b^8 \pi^2 r_+^7 F^{\prime}(r) \\
	\mathbf{A}^{\xi_2}(r) &= \frac{\pi^2 r_+^7 T^3 b^8}{r^2}\left(1-\frac{4 r_+ \mu^2}{3 r^3 \left(1+\frac{\mu^2}{6 r_+^2}\right)}\right)-\left(r^2 f(r){a^{\kappa}}^{\prime}(r)\right)^{\prime} \\
	\mathbf{A}^{\xi_3}(r) &= 0 \\
	\mathbf{A}^{\xi_4}(r) &= \frac{4 \sqrt{3} b^5 Q^3}{r^5}-\frac{2 \sqrt{3} b Q}{r^3} \\
	\mathbf{A}^{\tilde{\xi}_1}(r) &= -\sqrt{3}Q^2\left(\left(\frac{F(r)}{r^4}\right)^{\prime}+\frac{b^4}{2r^2}-b^4 F^{\prime}(r)\right)
\end{align}
\end{subequations}
and
\begin{subequations}
\label{E:Jsources}
\begin{align}
\notag
	\mathbf{J}^{\xi_1}(r) &=
		\frac{3\mu r^2}{8\pi^3 T^2}
		\left(\pi-2\arctan\left(\frac{r}{\pi T}\right)+\ln\left(\frac{r^2+\pi^2T^2}{(r+\pi T)^2}\right)\right) 
        \\
\label{E:Jxi1}
		&\phantom{=}-
		\frac{\left(3 r^5+4 \pi  T r^4+7 \pi ^2 T^2 r^3+10 \pi ^3 T^3 r^2+6 \pi ^4 T^4 r+4 \pi^5 T^5\right) \mu  r}{4 \pi \left(r^3+\pi  T r^2+\pi ^2 T^2 r+\pi ^3 T^3\right)^2}
		+\mathcal{O}\left(\mu^2\right) \\
	\mathbf{J}^{\xi_2}(r) &=r^2 \left(6 j^{\kappa}(r)+9 r {j^{\kappa}}^{\prime}(r)+r^2 {j^{\kappa}}^{\prime\prime}(r)\right) \\
	\mathbf{J}^{\xi_3}(r) &= -\frac{\left(r^3-r_+^3\right) b^3}{r^2 f(r)}\\
	\mathbf{J}^{\xi_4}(r) &= -2 b r \\
	\mathbf{J}^{\tilde{\xi}_1}(r) &= -\frac{1}{2}Q^3 b^4 \left(\frac{F^{\prime}(r)}{r}\right)^{\prime}.
\end{align}
\end{subequations}
The full expression for $\mathbf{J}^{\xi_1}(r)$ can be found in \eqref{E:Jxi1long}. The coefficients $\xi_i$ and $\tilde{\xi}_1$ in \eqref{E:xi1}-\eqref{E:xit1} were computed from \eqref{E:Asources} and \eqref{E:Jsources} with the help of \eqref{E:A2value} and \eqref{E:BtB}.


\section{Discussion}
\label{S:Discussion}
To summarize, on the gravity side our results show that one can extend the Reissner-Nordstr\"om AdS${}_5$ black hole solutions to black holes for which the charge density varies with the space-time coordinates.  The field theory dual of this configuration is a conformal fluid with a non vanishing chemical potential and non trivial $R$-charge current. The transport coefficients of this fluid are listed in \eqref{E:Transport0}, \eqref{E:Transport1}, \eqref{E:Transport2} and \eqref{E:Transportlong}.
Once the chemical potential vanishes one obtains the transport coefficients calculated in \cite{Bhattacharyya:2007jc,Baier:2007ix,Natsuume:2007ty} for the CFT dual of an uncharged black hole.

At first order in the derivative expansion we have found a transport coefficient of the $R$-charge current, $\Omega$, associated with the vorticity $\ell_{\mu}$ (see \eqref{E:defell} and \eqref{E:omega}) which, as far as we know, has not appeared in the literature so far. One can trace back the appearance of this component to the Chern-Simons term in \eqref{E:action}. If the Chern-Simons term were absent from the Lagrangian, $\Omega$ would vanish. A similar statement can be made for the second order transport coefficients $\tilde{\xi}_1$, $\tilde{\lambda}_1$ and $\tilde{\lambda}_2$.\footnote{Notice, however, that the presence of the Chern-Simons term is required by supersymmetry.}

Our result for the shear viscosity to entropy ratio agrees with \eqref{E:etaovers}. This observation has already been made in \cite{Son:2006em,Maeda:2006by,Mas:2006dy} and seems to be a feature of any gauge theory with a holographic dual. In light of the universality of $\eta/s$, it is natural to inquire if there are other hydrodynamic quantities whose value is universal \cite{Buchel:2007mf,Gubser:2007ni,Kovtun:2008kx}. In \cite{Kovtun:2008kx} it was suggested that the ratio of the electrical conductivity to the susceptibility of certain CFT's might have a universal value in the same sense as \eqref{E:etaovers}. In the $\mu \to 0$ limit
our result for $\kappa/\chi$ in \eqref{E:kappa} is consistent with the prediction of \cite{Kovtun:2008kx}. In that context it is also interesting to compare our full expression for $\kappa/\chi$ in \eqref{E:kappa} to the one obtained from the results of \cite{Son:2006em}, where the chemical potential of a different $U(1)$ subgroup of the $SO(6)$ $R$-symmetry group was turned on. As far as we can tell, the ratios are rather different. Also, the constant of proportionality in the analogue of the Wiedemann-Franz law \cite{LandP}, in our case
\begin{equation}
	\left(\frac{\frac{4}{3}\epsilon}{\rho T}\right)^2 \frac{\mu^2 \kappa}{\eta T} = 4 \pi^2,
\end{equation}
differs from the one obtained in \cite{Son:2006em} by a factor of two.

We have normalized all the second order transport coefficients \eqref{E:Transport2} relative to the second derivative of the energy density with respect to the temperature. This was done in order to conform to \cite{Haack:2008cp} and in order to get rid of some multiplicative constants related to the number of degrees of freedom of the gauge theory. Clearly, the expressions in \eqref{E:Transport2} are different from their $\mu=0$ counterparts. However, it is interesting to note that
\begin{equation}
	4 \lambda_1+\lambda_2=2\eta \tau_{\pi}
\end{equation}
for any value of $\mu$, and that this relation also holds, in the $\mu=0$ case, in more than d=4 transverse dimensions \cite{Haack:2008cp}. It would be nice to check if this relation remains valid even when $\mu \neq 0$ and $d>4$, or in other theories with a holographic dual.

Another interesting transport coefficient is $\lambda_3$ which vanishes when $\mu = 0$ both at strong and weak coupling \cite{Baier:2007ix}. When $\mu \neq 0$ it does not vanish, at least not at strong coupling. From a gravitational point of view, one reason for this difference is the Chern-Simons term in \eqref{E:action}. As discussed above, this term is responsible for the contributions proportional to $\ell_{\mu}$ in $j^{(1)}_\mu(r)$ and $a^{(1)}_\mu(r)$. These expressions contribute to the source $\mathbf{P}_{\mu\nu}^{\lambda_3}(r)$ via terms quadratic in $j^{(1)}_\mu(r)$ and $a^{(1)}_\mu(r)$.
One can verify that this is the origin of the second summand on the right hand side of \eqref{E:Plambda3}. However, even in the absence of the Chern-Simons term $\lambda_3$ would not vanish when $\mu \neq 0$. Of course, the choice of basis in \eqref{E:Secondorderterms} is not uniquely determined. it is always possible to redefine
\begin{equation}
\Sigma_5^{\prime} = \Sigma_5 + \frac{8 \mu \left(1+\frac{\mu^2}{6 r_+^2}\right)}{\pi^2 b^4 r_+^2 T^3}\Sigma_3
\end{equation}
so that the coefficient of $\Sigma_3$ vanishes both in the $\mu \neq 0$ case and in the $\mu \to 0$ limit. Further physical guidance would be needed in order to decide whether a certain choice of basis is preferred over another.

\vspace{2cm}


\noindent
{\bf \huge Appendix}

\begin{appendix}
\section{The vector modes' equations of motion}
\label{A:Vector}
In section \ref{S:Main} we have shown that the equations of motion for the perturbations of the metric and gauge field \eqref{E:EinsteinMaxwell} involve two vector modes $j_{\mu}^{(n)}(r)$ and $a_{\mu}^{(n)}(r)$ which are coupled,
\begin{align}
	\partial_r \left(r^3 f(r) \partial_r a(r)+2\sqrt{3} Q_0 j(r) \right) &= \mathbf{A}(r) \nonumber \\
	\partial_r \left(r^5\partial_r j(r)+2\sqrt{3} Q_0 a(r) \right) &= \mathbf{J}(r).\label{E:VectorEOMs}
\end{align}
We have removed the superscript $(n)$ specifying the order of the solution and the vector subscript $\mu$ for clarity.
In this section we show how to obtain the solution to these equations for arbitrary sources $\mathbf{A}(r)$ and $\mathbf{J}(r)$. The result was already
anticipated in \eqref{E:Vmodes}.

We start by considering the homogeneous version of these equations. The four homogeneous solutions are given by
\begin{subequations}
\label{E:SolutionHomV}
\begin{align}
	\begin{pmatrix}
	 	a(r) \\ j(r)
	\end{pmatrix}
	&=
	\begin{pmatrix}
		1 \\
		0
	\end{pmatrix},\quad
	\begin{pmatrix}
	 	a(r) \\ j(r)
	\end{pmatrix}
	=
	\begin{pmatrix}
		0 \\
		1
	\end{pmatrix},\quad
	\begin{pmatrix}
	 	a(r) \\ j(r)
	\end{pmatrix}
	=
	\begin{pmatrix}
		-\frac{\sqrt{3} Q_0 b_0^4}{r^2} \\
		\frac{1}{r^4}-\frac{b_0^4 Q_0^2}{r^6}
	\end{pmatrix}
	\equiv H_1(r)\ ,
\\
	\begin{pmatrix}
	 	a(r) \\ j(r)
	\end{pmatrix}
	&=
	\begin{pmatrix}
		-\frac{P_0}{r^2}+\frac{P_2(r)}{r^{2}}\sum_{x} \alpha(x)\ln|r^2-x^2| \\
		\frac14 b_0^4+\frac{P_6(r)}{r^6}
                + 2\sqrt{3} Q_0 f(r) \sum_x \alpha(x)\ln|r^2-x^2|
	\end{pmatrix}
	\equiv H_2(r)\ ,
\end{align}
\end{subequations}
where we have introduced the following notation:
\begin{align}
\notag
	P_0 &= \frac{27 \sqrt{3} b_0^{16}Q_0^5}{4\left(-4+27 b_0^{12}Q_0^4\right)},&
        P_2(r) &= -\frac{4}{b_0^4}\left(r^2-\frac{3}{2} b_0^4 Q_0^2\right),
	\\
        r^{-6} P_6(r) &= \frac{r^2 - 2\sqrt{3} P_0 Q_0 \poly}{P_2(r)}, &
	\alpha(x) &= \frac{\sqrt{3} b_0^{12}Q_0 (x^2+3 b_0^4 Q_0^2)}{8 \left(-1+3 x^4 b_0^4\right)\left(-4+27 b_0^{12}Q_0^4\right)}
\end{align}
and $x$ runs over the roots of \poly~which we denote by $\pm r_+$, $\pm r_-$ and $\pm i \sqrt{r_+^2+r_-^2}$ with $r_+>r_->0$. The outer horizon of the unperturbed black hole is located at $r=r_+$.
The two constant modes correspond to a deformation of the boundary, $H_1$ is a zero momentum quasi-normal mode corresponding to a shift in the boost parameters of the black brane, and $H_2$ is a solution which diverges at the horizons. In particular, at the outer horizon we find
\begin{equation}
\label{E:H2horizon}
	\lim_{r \to r_+}
	H_2(r) = \begin{pmatrix}
		\frac{\sqrt{3} Q_0}{2(2\pi T r_+^2)^2} \ln|r-r_+|+\mathcal{O}(r_+^0) \\
		\mathcal{O}(r_+^0)
	\end{pmatrix}.
\end{equation}
The rest of the solutions are finite at $r=r_+$.

To solve the non-homogeneous equations we integrate \eqref{E:VectorEOMs} once,
\begin{subequations}
\label{E:Firstorder}
\begin{align}
	\partial_r a(r)+2\sqrt{3}L Q_0 \frac{j(r)}{r^3 \poly}&=  \frac{\int^r \mathbf{A}(x)dx}{r^3 \poly}\\
	\partial_r j(r)+2\sqrt{3}L^3 Q_0 r^{-5} a(r) &= r^{-5}\int^r \mathbf{J}(x)dx.
\end{align}
\end{subequations}
For the moment we keep the lower limits of integration unspecified. The solutions to the homogeneous version of \eqref{E:Firstorder} which are first order equations, can be obtained from linear combinations of \eqref{E:SolutionHomV}. They are given by the columns of
\begin{equation}
\label{E:Hdef}
	H(r) =
	\begin{pmatrix}
		\frac{P_2(r)}{r^2} &
		-\frac{P_0}{r^2}+\frac{P_2(r)}{r^{2}}\sum_{x} \alpha(x)\ln|r^2-x^2| \\
		2\sqrt{3} Q_0 \poly &
		\frac{P_6(r)}{r^6}
                + 2\sqrt{3} Q_0 f(r) \sum_x \alpha(x)\ln|r^2-x^2|
	\end{pmatrix}\ .
\end{equation}
The overall multiplicative factor of the homogeneous solutions has been chosen so that $|H|=1$.

With the homogeneous
solutions at hand, one can use the method of variation of parameters to find
a particular solution to the inhomogeneous first order equations
\eqref{E:Firstorder}. This is given by
\begin{equation}
\label{E:solV1}
	\begin{pmatrix}
	 	a(r) \\ j(r)
	\end{pmatrix}
		=
	-H(r) \int_r
	 H^{-1}(x)
	\begin{pmatrix}
	 	\frac{\int^x \mathbf{A}(x^{\prime})dx'}{x^3 f(x)} \\
		x^{-5}\int^x \mathbf{J}(x^{\prime})dx'
	\end{pmatrix}
	dx\ .
\end{equation}

We choose the integration constants by requiring that the metric is differentiable up to and including the outer horizon and that there is no deformation of the boundary metric. Since $H$ is finite at the boundary, the latter requirement implies that we should set the upper limit of the outer integral to infinity. We assume this integral exists, i.e.,
\begin{equation}
\label{E:JAlarger}
	\mathbf{J}^{(n)}(r) = \mathcal{O}(r^2),\quad
	\mathbf{A}^{(n)}(r) = \mathcal{O}(r^0).
\end{equation}
If \eqref{E:JAlarger} does not hold then there is no asymptotically AdS solution. As stated in the main text, we have checked that \eqref{E:JAlarger} is satisfied up to second order in the derivative expansion. The other requirement, that the metric is differentiable, implies that we should set the lower limit of the inner integral to $r_+$ and add an appropriate multiple of the homogeneous solution $H_2$. Indeed, once the lower limit of the inner integral is set to $r=r_+$ then the outer integrand in \eqref{E:solV1} will be finite at the horizon. Thus, the only terms which may diverge at the horizon can arise from the logarithmic divergence in $H$ multiplying the outer integral in \eqref{E:solV1}. Since
\begin{multline}
\label{E:Dterm}
	\lim_{r\to r_+}
	\left[ -H(r) \int_r^{\infty}
	 H^{-1}(x)
	\begin{pmatrix}
	 	\frac{\int_{r_+}^x \mathbf{A}(x^{\prime})dx^{\prime}}{x^3 f(x)} \\
		x^{-5}\int_{r_+}^x \mathbf{J}(x^{\prime})dx^{\prime}
	\end{pmatrix}
	dx \right]
	=\\
	\begin{pmatrix}
        1\\0
        \end{pmatrix} \times
	\frac{\sqrt{3} Q_0}{2(2\pi T r_+^2)^2} \ln|r-r_+|
	\int_{r_+}^{\infty}
	\left(
	\frac{2 \sqrt{3} Q_0}{x^3}\int_{r^+}^x \mathbf{A}(x^{\prime})dx^{\prime}
	+
	\frac{1}{x^5}\left(\frac{4}{b_0^4} - \frac{6 Q_0^2}{x^2}\right)
	\int_{r_+}^x \mathbf{J}(x^{\prime})dx^{\prime}\right) dx
        \\+\mathcal{O}(r_+^0),
\end{multline}
then according to \eqref{E:H2horizon} one can get rid of the remaining logarithmic divergence by adding a term proportional to $H_2(r)$  to our solution. We are still left with one integration constant: the homogeneous solution $H_1$ neither deforms the boundary nor diverges at the horizon, so we may add it to \eqref{E:solV1} without spoiling the boundary conditions. As we have mentioned earlier, $H_1$ is the homogeneous solution associated with a shift in the boost parameters. From the point of view of the boundary theory, this corresponds to an ambiguity in the definition of the velocity field which is fixed by going to the Landau frame. Fixing the Landau gauge in the boundary theory corresponds to setting the fourth order coefficient of a near boundary expansion of $j^{(n)}(r)$ to zero, cf.\ \eqref{E:Landaugauge}. This precisely fixes the remaining integration constant. Our final result  for the solution to \eqref{E:VectorEOMs} is then
\begin{equation}
\label{E:solV2}
	\begin{pmatrix}
	 	a(r) \\ j(r)
	\end{pmatrix}
		=
	-H(r) \int_r^{\infty}
	 H^{-1}(x)
	\begin{pmatrix}
	 	\frac{\int^x_{r_+} \mathbf{A}(x^{\prime})dx'}{x^3 f(x)} \\
		x^{-5}\int^x_{r_+} \mathbf{J}(x^{\prime})dx'
	\end{pmatrix}
	dx
	+
	\coefone H_1(r)+\coeftwo H_2(r)\ ,
\end{equation}
where \coefone~is set to
\begin{multline}
\label{E:alpha1def}
	\coefone =-\lim_{r\to\infty}
    \Bigg[ \frac{1}{4}\left(\sum_{m=0}^2 \frac{(-1)^{m}r^{m+1}}{(m+1)!}\partial_r^m \mathbf{J}(r)-\int_{r_+}^r \mathbf{J}(x)dx\right)
    \\+\frac{9\sqrt{3}Q_0}{4 r^2}
    \left(\sum_{m=0,1} \frac{(-1)^m r^{m+1}}{3^{2m+1}}\partial_r^m \mathbf{A}(r)-\int_{r^+}^r \mathbf{A}(x)dx \right) \Bigg]
\end{multline}
by requiring that $j^{(\overline{4})}=0$ and
\begin{equation}
\label{E:alpha2def}
	\coeftwo = - \int_{r_+}^{\infty}
	\left(
	\frac{2 \sqrt{3} Q_0}{x^3}\int_{r^+}^x \mathbf{A}(x^{\prime})dx^{\prime}
	+
	\frac{1}{x^5}\left(\frac{4}{b_0^4} - \frac{6 Q_0^2}{x^2}\right)
	\int_{r_+}^x \mathbf{J}(x^{\prime})dx^{\prime}\right) dx
\end{equation}
by requiring that
the metric is differentiable at the outer
horizon (notice that this does not interfere with demanding
$j^{(\overline{4})}=0$ since
$H_2$ does not contain any term proportional to $r^{-4}$).
We point out that since $\mathbf{A} = \mathcal{O}(r^0)$, c.f. \eqref{E:JAlarger}, the second line in \eqref{E:alpha1def} will always evaluate to zero.
Also, in practice it is efficient to replace the double integrals in \eqref{E:alpha2def} with single integrals. This can be done by integrating by parts. Using \eqref{E:JAlarger},
we obtain
\begin{equation}
	\coeftwo = -
	\sqrt{3} Q_0\int_{r^+}^{\infty} \frac{\mathbf{A}(x)}{x^2}dx
	-
	\frac{1}{b_0^4}\int_{r^+}^{\infty} \frac{\mathbf{J}(x)}{x^4}dx
	+Q_0^2 \int_{r^+}^{\infty}\frac{\mathbf{J}(x)}{x^6}dx\ .
\end{equation}

Equation \eqref{E:A2value} in the main text was obtained by expanding \eqref{E:solV2} in a series expansion near
the boundary, using \eqref{E:JAlarger} and extending the solution from the neighborhood of $x_0^{\mu}$ to $\mathbf{R}^{3,1}$.

\section{Long expressions}
\label{A:Long}
In certain places in the main text the expressions we have found were somewhat long. In this appendix we have collected the expressions which were omitted.

The full expressions for the transport coefficients $\lambda_4$ and $\xi_1$ whose expansion (in $\tfrac{\mu}{T}$) appeared in section \ref{S:Conformalfd}, equations \eqref{E:lambda4} and \eqref{E:xi1},
are given by
\begin{subequations}
\label{E:Transportlong}
{\tiny
\begin{multline}
\frac{\lambda_4}{c} = \frac{\left(9 r_-^{16}+36 r_+^2 r_-^{14}+372 r_+^4 r_-^{12}+990 r_+^6 r_-^{10}+1523 r_+^8 r_-^8+1438
   r_+^{10} r_-^6+696 r_+^{12} r_-^4+136 r_+^{14} r_-^2-16 r_+^{16}\right) T^4}{3456 r_+^4 \left(1+\frac{\mu ^2}{6 r_+^2}\right)^2 \left(2
   r_-^2+r_+^2\right)^2 \left(r_+^4 + r_+^2 r_-^2 +r_-^4 \right)^3}\\
-
   \frac{\left(r_-^4+r_+^2 r_-^2-2 r_+^4\right)^6 \left(1+\frac{\mu^2}{6 r_+^2}\right) 
\ln \left(\frac{r_+^2-r_-^2}{r_-^2+2 r_+^2}\right)}{1152 \pi ^4 r_+^6 \left(2 r_-^2+r_+^2\right)^3 \left(r_-^4+r_+^2 r_-^2+r_+^4\right)
   \left(r_-^4+r_+^2 r_-^2+2 r_+^4\right)^2}
\end{multline}
}
and
\begin{multline}
\label{E:xi1long}
	\frac{8\pi^2}{N^2} \xi_1 =
	\frac{r_+ \left(r_-^2+r_+^2\right) \left(13 r_-^4+13 r_+^2 r_-^2+10 r_+^4\right) \left(r_-^5+r_+^2 r_-^3-2 r_+^4
   r_-\right)^2}{32 \pi  \left(r_-^4+r_+^2 r_-^2+r_+^4\right)^3 \left(r_-^4+r_+^2 r_-^2+2 r_+^4\right)}\\
	-
	\frac{\left(r_-^4+r_+^2 r_-^2-2 r_+^4\right)^3}{8 \pi  r_+ \left(r_-^4+r_+^2 r_-^2+r_+^4\right) \left(2 r_-^6+3
   r_+^2 r_-^4+5 r_+^4 r_-^2+2 r_+^6\right)}\ln\left(\frac{r_+^2-r_-^2}{2 r_+^2+r_-^2}\right)\ .
\end{multline}
\end{subequations}

The source terms missing from  \eqref{E:Plambda4} and \eqref{E:Jxi1} in section \ref{S:Main} are
\begin{subequations}
\label{E:Longsources}
{\tiny
\begin{multline}
\label{E:Plambda4long}
 	\mathbf{P}^{\lambda_4}=
		-\frac{\sqrt{r_-^2+r_+^2} \left(r_-^4+r_+^2 r_-^2-2 r_+^4\right)^3}{12 \sqrt{3} \pi  r_- r_+^2 \left(r_-^{10}+3 r_+^2
   r_-^8+6 r_+^4 r_-^6+7 r_+^6 r_-^4+5 r_+^8 r_-^2+2 r_+^{10}\right)}
		\left(r^3 j^{\kappa(\partial\beta)}(r)\right)^{\prime}\\
		+
		\frac{\sqrt{r_-^2+r_+^2} \left(r_-^4+r_+^2 r_-^2-2 r_+^4\right)^3}{12 \sqrt{3} \pi  r_- r_+^2 \left(r_-^{10}+3 r_+^2
   r_-^8+6 r_+^4 r_-^6+7 r_+^6 r_-^4+5 r_+^8 r_-^2+2 r_+^{10}\right)}
	\left(r^3 j^{\kappa(r)}\right)^{\prime}\\
		-\frac{5 \left(r_-^4+r_+^2 r_-^2-2 r_+^4\right)^3}{12 \pi  r_+ \left(r_-^8+2 r_+^2 r_-^6+4 r_+^4 r_-^4+3 r_+^6
   r_-^2+2 r_+^8\right)}
		\left(\frac{a^{\kappa}(r)}{r}\right)^{\prime}
		+2 \sqrt{3} r_- r_+ \sqrt{r_-^2+r_+^2} \left(a^{\kappa}(r)j^{\kappa}(r)\right)^{\prime}
		-\frac{\left(r_-^4+r_+^2 r_-^2-2 r_+^4\right)^3}{8 \pi  r_+^2 \left(r_-^4+r_+^2 r_-^2+r_+^4\right)^2}{a^{\kappa}}^{\prime}(r)\\
		+
		\frac{\sqrt{3} \left(r_-^4+r_+^2 r_-^2-2 r_+^4\right)^3}{4 \pi  r_+ \left(r_-^8+2 r_+^2 r_-^6+4 r_+^4 r_-^4+3 r_+^6
   r_-^2+2 r_+^8\right)}
		\left(r^3 j^{\kappa(\partial Q)}(r)\right)^{\prime}
		-\frac{\left(r_-^4+r_+^2 r_-^2-2 r_+^4\right)^3}{4 \pi  r_+ \left(r_-^8+2 r_+^2 r_-^6+4 r_+^4 r_-^4+3 r_+^6
   r_-^2+2 r_+^8\right)}\frac{a^{\kappa}(r)}{r^2}
\end{multline}
}
and
{\small
\begin{multline}
\label{E:Jxi1long}
\mathbf{J}^{\xi_1}(r) =\bigg(
            \left(r^5 F^{\prime}(r)j^{\kappa}(r)\right)^{\prime}\\
            -\frac{\pi  \sqrt{r_+^2+r_-^2} r_- (r-r_+)^2 r_+^6 \left(3 r_+ \left(r^2+2 r_+ r+3 r_+^2\right) r^2+(2 r+r_+) \left(r_-^4+r_+^2
   r_-^2+4 r_+^4\right)\right) T^2}{\sqrt{3} r^8 \left(r_-^8+2 r_+^2 r_-^6+4 r_+^4 r_-^4+3 r_+^6 r_-^2+2 r_+^8\right)
   f(r)^2} \bigg)\ ,
\end{multline}
}
\end{subequations}
where we have defined
\begin{equation}
	\partial_{\mu}j^{\kappa}(r) = j^{\kappa(\partial\beta)}(r) u^{\alpha}\partial_{\alpha}u_{\mu}
		+ j^{\kappa(\partial Q)}(r)\partial_{\mu}Q\ .
\end{equation}

Finally, in \eqref{E:j1a1} we have parameterized the first order bulk solutions for the vector modes by two functions $a^{\kappa}(r)$ and $j^{\kappa}(r)$. The first, $a^{\kappa}(r)$, is given by
{\tiny
\begin{multline}
\label{E:akappalong}
	a^{\kappa}(r)=
	\frac{\sqrt{r_-^2+r_+^2} \left(2 r_-^4+3 r_+^2 r_-^2+2 r_+^4\right) \left(r_+^2-r_-^2\right)^3}{4 r_+ \left(2
   r_-^2+r_+^2\right)^3 \left(r_-^4+r_+^2 r_-^2+2 r_+^4\right)}
	-\frac{3 r_-^2 r_+ \left(r_-^6+2 r_+^2
   r_-^4-r_+^4 r_-^2-2 r_+^6\right)}{2 \pi  r \left(2 r_-^2+r_+^2\right)^2 \left(r_-^4+r_+^2 r_-^2+2
   r_+^4\right)}\\
	+\frac{27 r_+^2 \left(r_-^2+r_+^2\right)^2 \left(r_-^4+r_+^2 r_-^2-2 r_+^4\right) r_-^4}{16 r^2 \pi  \left(2
   r_-^6+3 r_+^2 r_-^4+3 r_+^4 r_-^2+r_+^6\right)^2}+\frac{3 r_+ \left(r_-^2-r_+^2\right)^3 \sqrt{r_-^2+r_+^2} \left(2
   r_-^6+5 r_+^2 r_-^4+5 r_+^4 r_-^2+2 r_+^6\right) r_-^2}{8 r^2 \left(2 r_-^2+r_+^2\right)^3 \left(r_-^8+2 r_+^2
   r_-^6+4 r_+^4 r_-^4+3 r_+^6 r_-^2+2 r_+^8\right)}\\
		+\left(\frac{\left(r_-^2-r_+^2\right)^3 \left(2 r_-^6+5 r_+^2
   r_-^4+5 r_+^4 r_-^2+2 r_+^6\right)}{2 \pi  r_+ \sqrt{r_-^2+r_+^2} \left(2 r_-^2+r_+^2\right)^3 \left(r_-^4+r_+^2
   r_-^2+2 r_+^4\right)}-\frac{3 r_-^2 r_+ \left(r_-^2-r_+^2\right)^3 \left(r_-^2+r_+^2\right)^{3/2} \left(2 r_-^4+3 r_+^2
   r_-^2+2 r_+^4\right)}{4 \pi  r^2 \left(2 r_-^2+r_+^2\right)^3 \left(r_-^8+2 r_+^2 r_-^6+4 r_+^4 r_-^4+3 r_+^6
   r_-^2+2 r_+^8\right)}\right) \arctan\left(\frac{r}{\sqrt{r_-^2+r_+^2}}\right)\\
		+\left(\frac{r_- \left(r_-^2+r_+
   r_-+r_+^2\right) \left(r_-^3-r_+ r_-^2+2 r_+^2 r_--2 r_+^3\right)^3}{4 \pi  r_+^2 \left(2 r_-^2+r_+^2\right)^3
   \left(r_-^4+r_+^2 r_-^2+2 r_+^4\right)}-\frac{3 r_-^3 \left(r_-^2+r_+^2\right) \left(r_-^3-r_+ r_-^2+2 r_+^2
   r_--2 r_+^3\right)^3}{8 \pi  r^2 \left(2 r_-^2+r_+^2\right)^3 \left(r_-^6-r_+ r_-^5+2 r_+^2 r_-^4-r_+^3 r_-^3+3
   r_+^4 r_-^2-2 r_+^5 r_-+2 r_+^6\right)}\right) \ln (r-r_-)\\
		+\left(\frac{r_- \left(r_-^2-r_+ r_-+r_+^2\right)
   \left(r_-^3+r_+ r_-^2+2 r_+^2 r_-+2 r_+^3\right)^3}{4 \pi  r_+^2 \left(2 r_-^2+r_+^2\right)^3 \left(r_-^4+r_+^2
   r_-^2+2 r_+^4\right)}-\frac{3 r_-^3 \left(r_-^2+r_+^2\right) \left(r_-^3+r_+ r_-^2+2 r_+^2 r_-+2 r_+^3\right)^3}{8
   \pi  r^2 \left(2 r_-^2+r_+^2\right)^3 \left(r_-^6+r_+ r_-^5+2 r_+^2 r_-^4+r_+^3 r_-^3+3 r_+^4 r_-^2+2 r_+^5
   r_-+2 r_+^6\right)}\right) \ln (r+r_-)\\
		+\left(\frac{3 r_-^2 r_+^2 \left(r_-^2+r_+^2\right)}{4 \pi  r^2 \left(r_-^4+r_+^2
   r_-^2+r_+^4\right)}-\frac{1}{2 \pi }\right) \ln (r+r_+)
		+\left(\frac{3 r_-^2 \left(r_-^2+r_+^2\right)^2 \left(r_-^2-r_+^2\right)^3}{8
   \pi  r^2 \left(2 r_-^2+r_+^2\right)^3 \left(r_-^4+r_+^2 r_-^2+r_+^4\right)}+\frac{\left(r_+^2-r_-^2\right)^3
   \left(r_-^2+r_+^2\right)}{4 \pi  r_+^2 \left(2 r_-^2+r_+^2\right)^3}\right) \ln \left(r^2+r_-^2+r_+^2\right)
\end{multline}
}
and $j^{\kappa}(r)$ can be determined from $a^{\kappa}(r)$ through the equation of motion,
\begin{multline}
\label{E:jkappalong}
 	j^{\kappa}(r)=
	\frac{\left(r_-^4+r_+^2 r_-^2-2 r_+^4\right)^3 \left(r \left(r_-^4+r_+^2 r_-^2+2 r_+^4\right)-2 r_+ \left(r_-^4+r_+^2
   r_-^2+r_+^4\right)\right)}{16 \sqrt{3} \pi  r r_- r_+^3 \sqrt{r_-^2+r_+^2} \left(r_-^4+r_+^2 r_-^2+r_+^4\right)^2
   \left(r_-^4+r_+^2 r_-^2+2 r_+^4\right)}\\
	-\frac{\left(r_-^2-r^2\right) \left(r_+^2-r^2\right) \left(r^2+r_-^2+r_+^2\right)}
   {2 \sqrt{3} r^3 r_- r_+ \sqrt{r_-^2+r_+^2}}{a^{\kappa}}'(r).
\end{multline}

\end{appendix}


\section*{Acknowledgments}
We would like to thank N.~Banerjee, J.~Bhattacharya, S.~Bhattacharyya, G.~Cardoso, S.~Dutta, A.~Karch, R.~Loganayagam, D.~Son, Y.~Stanev, A.~Starinets, P.~Sur\'owka, D.~Tsimpis and M.~Zagermann for useful discussions.
This work is supported
in part by the European Community's Human Potential Program under
contract MRTN-CT-2004-005104 ``Constituents,
fundamental forces and symmetries of the universe'' and the Excellence Cluster
``The Origin and the Structure of the Universe'' in Munich.
M.H.~is supported by the German
Research Foundation (DFG) within the Emmy-Noether-Program (grant number:
HA 3448/3-1). A.Y.~is supported in part by the German Research Foundation and by the Minerva foundation.

\bibliographystyle{JHEP}
\bibliography{HigherO}

\end{document}